\documentclass[journal]{IEEEtran}

\usepackage{cite}      

\usepackage{graphicx}
\usepackage{amsmath}
\usepackage{amsfonts}
\usepackage{url}
\usepackage{color}
\usepackage[normalem]{ulem}
\usepackage{mathabx,graphicx}
\usepackage{epstopdf}

\makeatletter
\@ifpackageloaded{biblatex}{\addbibresource{literatur.bib}}{\bibliography{literatur}}
\makeatother

\newcommand{\be}{\begin{equation}}
\newcommand{\ee}{\end{equation}}
\newcommand{\bea}{\begin{eqnarray}}
\newcommand{\eea}{\end{eqnarray}}
\newcommand{\beaa}{\begin{eqnarray*}}
\newcommand{\eeaa}{\end{eqnarray*}}
\newcommand{\ben}{\begin{enumerate}}
\newcommand{\een}{\end{enumerate}}
\newcommand{\bi}{\begin{itemize}}
\newcommand{\ei}{\end{itemize}}

\begin{document}

\title{Kramers Kronig PAM transceiver and two-sided polarization-multiplexed Kramers Kronig transceiver}


\author{Cristian Antonelli, Antonio Mecozzi, and Mark Shtaif
\thanks{C. Antonelli and A. Mecozzi are with the Department of Physical and Chemical Sciences,
University of L'Aquila, L'Aquila 67100, Italy. M. Shtaif is with the Department of Physical Electronics,
Tel Aviv University, Tel Aviv 69978, Israel.}}


\maketitle

{ \begin{abstract}
We propose two transceiver schemes based on Kramers Kronig (KK) detection. One targets low-cost high-throughput applications and uses PAM transmission in combination with direct detection and digital reconstruction of the optical phase. This scheme allows digital compensation of chromatic dispersion and provides a significant improvement in terms of spectral efficiency, compared to conventional PAM transmission. The second scheme targets high-channel-count coherent systems with the aim of simplifying the receiver complexity by reducing the optical components count.
This scheme is based on the transmission of two SSB signals that share a transmission laser, each obtained by suppressing either optically or digitally one side of a real-valued signal. At the receiver, the two SSB signals are received separately, after optical de-interleaving, by KK detection, using a single local oscillator laser.
\end{abstract}}


\section{Introduction}\label{Introduction}
The simplification of optical communication receivers for short and intermediate range systems has become one of the most important problems in fiber communications in the past few years. The challenge is to reduce the cost of the transceiver with as little compromise on its performance and throughput as possible. Currently, it appears that pulse amplitude modulation (PAM) \cite{Eiselt} is leading the race for low-cost transmission systems, although significant competition has been presented by other direct detection techniques, such as the ones reported in \cite{Randel_invited,Lowery,Schmidt,Petermann,Randel}. Recently, we have proposed a new receiver scheme \cite{KKOptica} that allows the reconstruction of complex constellations from an intensity measurement requiring a single photo-diode. The method proposed in \cite{KKOptica} relies on transmitting together with the information signal, a CW tone positioned slightly outside of the signal'™s spectrum. We have shown that, as soon as the intensity of the CW tone exceeds by a few decibels the power of the information-carrying signal \cite{KKOptica}, the overall electric field becomes minimum phase, implying that its complex envelope can be extracted from an intensity measurement. The implementation of the KK receiver has been tested in the last year with a number of experimental realizations \cite{KKOFCpdp,KKJLT,KKECOC,LiECOC17a,LiECOC17b,ECOC17Plant,Stuttgart,PDPIEEEPC}.

Another critical challenge that is being faced in the arena of long-haul communications is the reduction of the receiver spatial footprint in systems where a large multiplicity of frequency and possibly spatial channels \cite{KDDI} are received simultaneously. In these cases, the capability of accommodating polarization multiplexing is an additional important requirement. While typical coherent receivers can accommodate the signal multiplicity, it is quite clear that they cannot comply with the spatial footprint requirement, as the number of received channels increases, unless special solutions are envisaged. In a recent paper \cite{KKPolMux}, we have shown that the KK transceiver can accommodate polarization multiplexing by adding the necessary CW tone at the receiver, provided that a local oscillator (LO) is available. This solution requires only two photo-diodes and no interferometric optics, versus eight photo-diodes and two optical hybrids required by an intradyne receiver. Clearly, in this situation the use of the KK approach provides  a substantial gain in terms of complexity and compactness, and in principle, it involves no performance penalty. The polarization-multiplexed KK transceiver is also a good candidate to reduce the spatial footprint of high-channel-count coherent receivers, and its use in a dense-WDM environment has been demonstrated recently in \cite{KKECOC}.

In this work we propose two new transceiver schemes based on the KK approach. The first scheme addresses the former challenge of reducing the transceiver cost and complexity in the context of short and intermediate range systems. This scheme was first presented in \cite{KKOFC} and is reviewed here. We call it the KK-PAM scheme, and it is based on separately modulating two field quadratures. One field quadrature contains a non-negative PAM signal, whereas the other contains its Hilbert transform, such that the overall field that is combined from the two quadratures is single-sideband (SSB). The generated SSB signal is also minimum phase, and hence it can be fully reconstructed from its intensity measurement. After the field is reconstructed, it is separated into its two original quadratures (as described later), and its in-phase quadrature is used to recover the transmitted data. 
The second scheme addresses more specifically the latter challenge of reducing the spatial footprint in receivers for medium and long-haul systems, while still addressing the receiver complexity issue. We call this scheme the two-sided (TS) KK transceiver scheme, and it combines the spectral efficiency of the KK-PAM transceiver with the ability to accommodate polarization multiplexing \cite{KKPolMux}. In the TS-KK scheme two SSB signals share the same transmit laser and LO, and each of them is obtained by suppressing either optically or digitally one side of a real-valued signal. At the receiver, the two SSB signals are received separately, after optical de-interleaving. In order to make this approach compatible with the optical filtering capabilities of WDM components that are available today, it is necessary to introduce a small frequency gap between the spectra of two SSB signals that form a WDM channel. 

The paper is organized as follows. In Section \ref{KKPAM} we review the KK-PAM transceiver introduced in \cite{KKOFC} and proceed to its numerical validation. In Section \ref{TSKK} we illustrate the principle of operation of the TS-KK scheme, and validate it by means of numerical simulations. Section \ref{Conclusions} is devoted to the conclusions.
 
\section{KK-PAM transceiver}\label{KKPAM}
The principle underpinning the KK-PAM transceiver is described in Fig. \ref{SchemePAM}. We consider an optical PAM signal 
\be p(t)=A+\sum_k a_k g(t-kT)\ee
where $a_k$ are zero-mean real-valued data symbols, and $g(t)$ is the fundamental symbol waveform, which is also assumed to be real valued, and its spectrum is symmetric and contained between  $-B/2$ and $B/2$ (for Nyquist pulses $T=1/B$). The parameter $A$ denotes a positive bias value that serves to ensure the non-negativity of $p(t)$. The  KK-PAM scheme relies on the transmission of the SSB version of $p(t)$ given by 
\be s(t)=A+ \sum_k a_k g_{\mathrm{SSB}} (t-kT), \ee
where $g_{\mathrm{SSB}} (t)=g(t)+iH\{g(t)\}$, with $H\{\cdot\}$ denoting the Hilbert transform. The spectrum of $g_{\mathrm{SSB}} (t)$ is contained between 0 and $B/2$. 

At the receiver, after filtering out-of-band noise, the optical signal is photo-detected, and the photo-current $I(t)$ is processed according to the KK reconstruction algorithm detailed in \cite{Mecozzi2016}. 
A key-step in this algorithm consists of recovering the phase $\phi(t)$ of the SSB signal $s(t)$ by means of the relation
\be \phi(t) = \frac i 2 H\{\log[I(t)]\}. \label{Phase}\ee
The reconstructed field is given by $\sqrt{I(t)} \exp\left[ i \phi(t) \right]$, and it is equal to the optical field impinging upon the receiver up to a constant (time independent) phase difference. Note that the entire process of field reconstruction is performed in the digital domain. This implies that the $I(t)$ needs to be sampled at least at its Nyquist bandwidth of $B$, and then digitally up-sampled so as to accommodate the bandwidth expansion implied by the logarithm of Eq. (\ref{Phase}). An up-sampling factor of the order of two or three ha been shown to be sufficient \cite{KKOptica,LiECOC17b}. The KK procedure recovers the dispersed version of the SSB signal $s(t)$, while preserving the phase of the information-carrying signal relative to the bias, thus making additional phase recovery unnecessary. After digital  CD compensation, $s(t)$ is recovered, and its real part yields the desired PAM signal. Since $\mathrm{Real}\{s(t)\} = p(t)$, the mere non-negativity of $p(t)$ ensures that $s(t)$ never encircles the origin in the complex plane, and hence it satisfies the minimum-phase condition underpinning the KK reconstruction procedure \cite{Mecozzi2016}.   Yet, in the presence of chromatic dispersion, the value of $A$ may have to be increased slightly in order to guarantee that the received optical signal satisfies the minimum-phase condition. These considerations are discussed in detail in Ref. \cite{Mecozzi2016}.  

\subsection{Numerical validation of the KK-PAM transceiver}
\begin{figure}
	\centering\includegraphics[width=0.95\columnwidth]{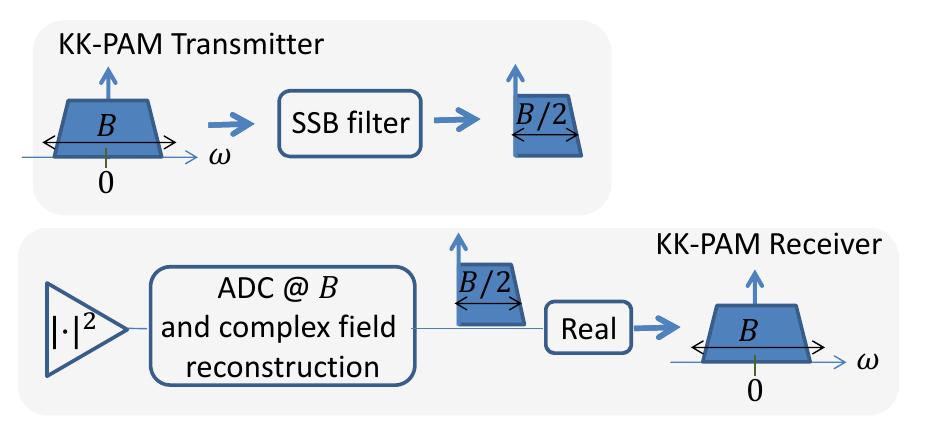}
	\caption{Principle of operation of the KK-PAM transceiver. At the transmitter, one side of a non-negative real-valued PAM signal of bandwidth $B$ is suppressed by means of a SSB filter (which can be implemented either in the digital domain or in the form of an  optical filter). At the receiver, the detected photocurrent is sampled with a sampling rate of $B$, and the SSB optical signal is reconstructed by applying the KK algorithm. The real PAM signal is obtained by taking the real part of the reconstructed signal. }\label{SchemePAM}
\end{figure}

\begin{figure}
	\centering\includegraphics[width=0.9\columnwidth]{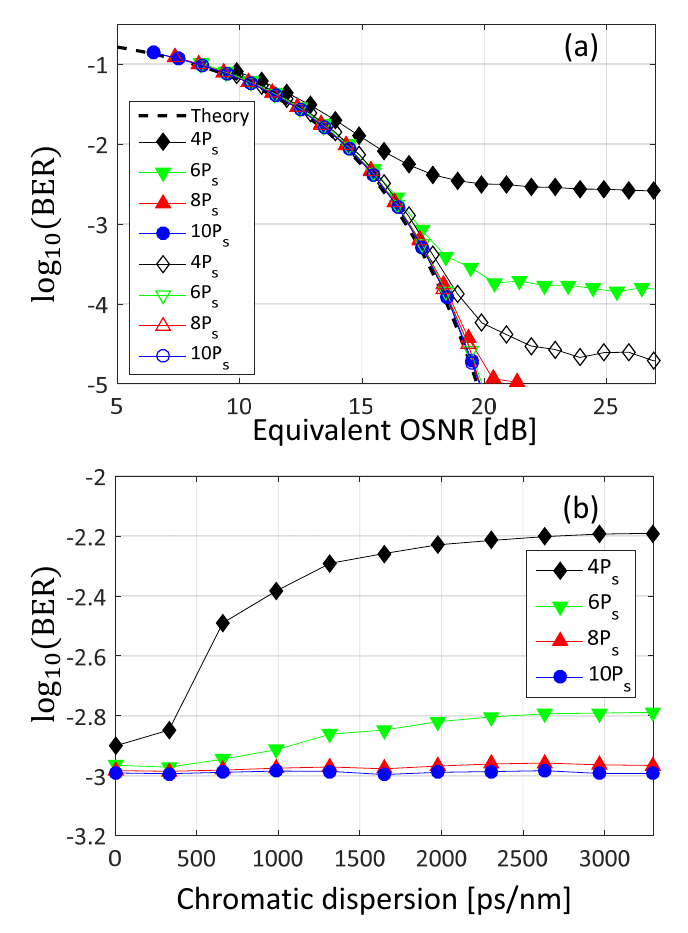}
	\caption{(a) The pre-FEC BER of a 48-Gbaud KK 4-PAM system (100-km SSMF), versus the equivalent OSNR for the displayed values of the CW bias power. Empty markers show the BER after optical CD compensation, whereas the filled markers show the BER when the CD is compensated for digitally. (b) The pre-FEC BER versus the link chromatic dispersion, for the same KK-PAM system. The curves were obtained by keeping the power of the information carrying signal fixed, and with digital CD compensation.}\label{BERpam1}
\end{figure}

\begin{figure}
	\centering\includegraphics[width=0.9\columnwidth]{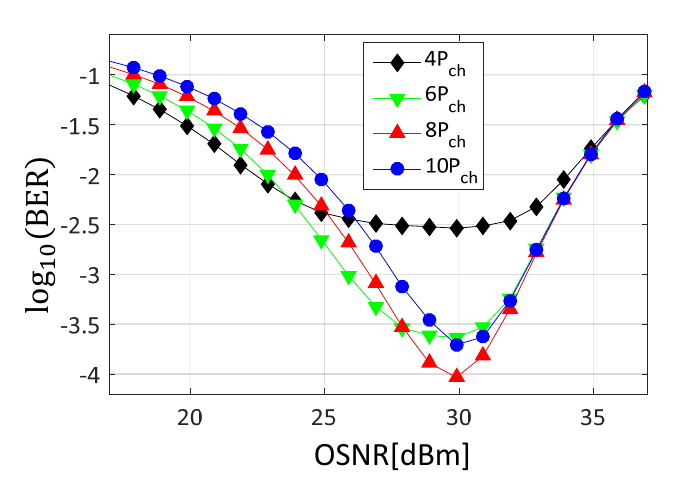}
	\caption{The BER of the central channel of a  WDM system with five 40-GHz spaced KK-PAM modulated channels, versus the equivalent OSNR (which is the OSNR that one would measure if there were no bias component in the transmitted signal). The dashed curve shows the theoretical BER of a 4-PAM coherent system impaired by AWGN \cite{Proakis}. }\label{BERpam2}
\end{figure}

In this section we numerically validate the KK-PAM transceiver described in the previous section. We consider a 100 km link over standard single-mode fiber (SSMF), where we transmit a 4-PAM modulated signal using a raised cosine fundamental waveform with a roll-off factor of 0.05, at a symbol rate of 48 Gbaud. This  translates into an optical bandwidth of approximately 24 GHz for the transmitted SSB signal. In the numerical results that we present in this section, the SSB signal $s(t)$ is assumed to have been produced by an I/Q modulator. The optical spectrum of $s(t)$ if shown in Fig. \ref{SchemePAM}. In the simulations, we assumed an overall loss budget of 26 dB and optical pre-amplification with a noise figure of 5 dB. We then evaluate the BER by transmitting a pseudo-random sequence of $2^{15}$ Gray-coded symbols. At the receiver, an optical filter with a 12th order super-Gaussian shape having a 3 dB bandwidth of 26 GHz, and centered at 16 GHz was used to remove the excess noise. In Fig. \ref{BERpam1}a we plot the BER as a function of the equivalent OSNR, defined as $\mathrm{{OSNR}_{eq}} = P_s/P_n$, which is the OSNR that one would measure in the absence of the bias component. Here $P_s$  denotes the power  contained in the zero-mean information carrying signal $\sum_k a_k g_{\mathrm{SSB}}(t-kT)$, and $P_n$ denotes the noise power within a bandwidth of 0.1 nm.
The various curves correspond to the different bias levels as shown in the legend, where the bias power $A^2$ is specified as a multiple of  $P_s$. The empty markers represent the case where CD is compensated optically prior to detection, whereas the filled markers represent the case where CD is compensated for digitally after signal reconstruction. The difference between the two cases follows from the fact that in the presence of dispersion, the non-negativity condition described earlier may be violated, in which case the field reconstruction is no longer perfect. This issue is notable particularly when the bias level is small, whereas in the case of a stronger bias the non-negativity is maintained even for the dispersed signal and the difference between the curves plotted with the filled and empty markers vanishes. The saturation of the BER curves in the absence of optical CD compensation when $A^2 = 4P_s$ and $A^2 = 6P_s$, follows from the fact that at high SNR levels the BER is dominated by imperfect field reconstruction and hence it does not improve with power. Naturally, no saturation of the BER occurs when CD is optically compensated (empty markers). All of the curves shown in Fig. \ref{BERpam1}a comply with an FEC threshold of $10^{-2}$. When considering the performance in terms of the equivalent OSNR, and under the assumption of perfect reconstruction, the KK-PAM receiver is equivalent to a coherent receiver, whose theoretical BER curve \cite{Proakis} is plotted by a dashed line in Fig. \ref{BERpam2}. 
In Fig. \ref{BERpam1}b we show the robustness of the KK-PAM scheme to CD, by plotting the BER as a function of total CD, while assuming an equivalent OSNR of $17$ dB. 
The CD tolerance clearly increases considerably with the bias level, although even in the case where the bias is equal to $4P_s$ the overall performance is compliant with common FEC requirements. Indeed, moderate CD, as high as approximately 300 ps/nm, can be compensated digitally with no penalty in BER even at the lowest reported bias value of $4P_s$. When the bias level is increased to $8P_s$ or $10P_s$'  any value of CD can be compensated digitally with no penalty. In Fig. \ref{BERpam2} we investigate the performance of the KK-PAM scheme in the nonlinear regime, by repeating the simulations described in the context of Fig. \ref{BERpam1}a, with 5 WDM channels separated by 40 GHz and without optical CD compensation. The BER of the central channel is plotted as a function of the true (not the equivalent) OSNR. 
The effect of the nonlinearity is seen in the form of a rise of the BER curve. Only in the case of low-bias-power $A^2 = 4P_s$ reconstruction errors precede the effect of the nonlinear distortion. 


\subsection{Discussion} 
The KK-PAM scheme should not be misinterpreted as regular self-heterodyne or standard SSB transmissions, with the bias signal playing the role of a carrier. With these methods, the required carrier amplitude would have to be much larger than it is in the KK-PAM scheme.

The availability of the complex signal at the end of the reconstruction makes the KK-PAM scheme equivalent to a coherent receiver. One obvious consequence of this equivalence is the possibility of digitally compensating not only for CD, but also for other propagation effects. Another important implication is the fact that the capacity of this scheme is close to that of a scheme using a coherent receiver, because of the ability of the KK receiver to detect the 
amplitude and phase of the signal \cite{AntonioMarkJLT}. 

An obvious disadvantage of the scheme described in this section is that it requires a relatively expensive I/Q modulator, which is an undesirable constraint in the case of single-span low-cost systems. This disadvantage can be remedied to some extent by using a single-ended Mach-Zehnder modulator driven with two voltages \cite{MunichChinese}, although the potential of this solution for the implementation of the KK scheme is yet to be evaluated. Another option for generating a SSB signal of the kind required by the KK procedure, is to combine an amplitude modulator with an optical filter. However, the requirements on the optical filter are very critical, and hence its implementation is not straightforward. This option is investigated in the section that follows, where we present a more spectrally efficient KK scheme that accommodates for polarization multiplexing.

\section{Two-sided Polarization-multiplexed KK Transceiver}\label{TSKK}
In this section we propose a new KK scheme inspired by the KK-PAM transceiver that accommodates polarization-multiplexed transmission. In order to achieve this, we assume that the CW reference tone is available as a LO at the receiving edge, and hence needs not be transmitted with the information-carrying signal. This assumption is well justified, provided that a LO can be extracted from the transmission laser that is co-located with the receiver that performs the KK processing. In such cases the cost implications of using a LO are minimal. 
\begin{figure}
	\centering\includegraphics[width=0.9\columnwidth]{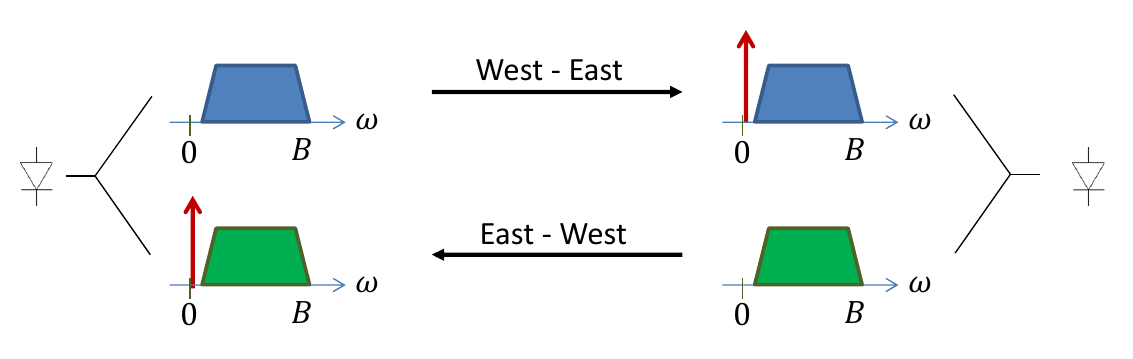}
	\caption{Original implementation of the KK scheme \cite{KKOptica}. The west and east transmitters generate SSB signals, whose bandwidth is contained between 0 and $B$, where the laser frequency is 0. The west-side laser is used both for generating the SSB signal going east and as a LO for receiving the signal arriving from the east location and vice-versa.}\label{KKbasic}
\end{figure}
In the context of the KK scheme, the most straightforward implementation is illustrated in Fig. \ref{KKbasic}. The west and east transmitters generate SSB signals, whose bandwidth is contained between 0 and $B$, where the laser frequency is 0. The west-side laser is used both for generating the SSB signal going east and as a LO for receiving the signal arriving from the east location. Similarly, the east-side laser also serves for the two functionalities. 
\begin{figure}
	\centering\includegraphics[width=0.9\columnwidth]{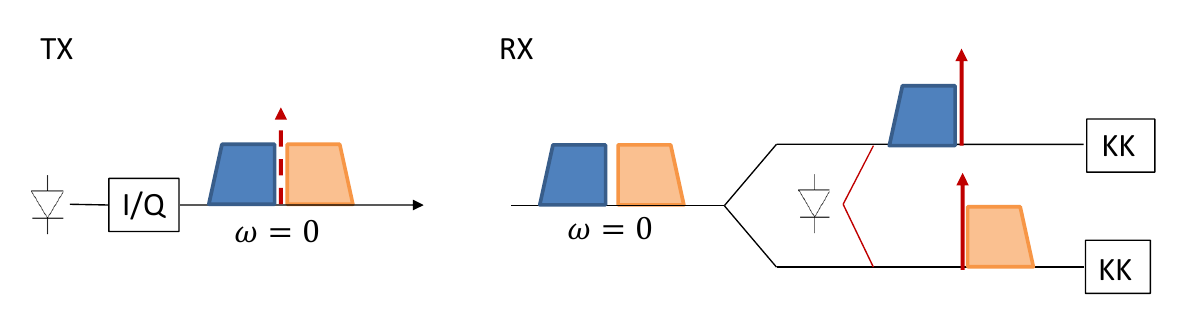}
	\caption{Operating principle of the TS-KK transceiver. The modulator generates a two-sideband spectrum, where each sideband contains an independent data-carrying signal. A guard-band around the center frequency ($\omega=0$) is inserted so as to allow separation between the two sidebands at the receiver, where the two sidebands are optically separated and KK-processed independently and in parallel. The two reception processes make use of the same LO at $\omega=0$.}\label{OperPrinc}
\end{figure}
Note that in this implementation the bandwidth of both the transmitter and receiver must be equal to $2B$, namely twice the bandwidth of the transmitted optical signal. A modified scheme, where the bandwidth of the transmitter is reduced to $B$ has been recently  demonstrated in \cite{KKECOC}. 
\begin{figure}
	\centering\includegraphics[width=0.9\columnwidth]{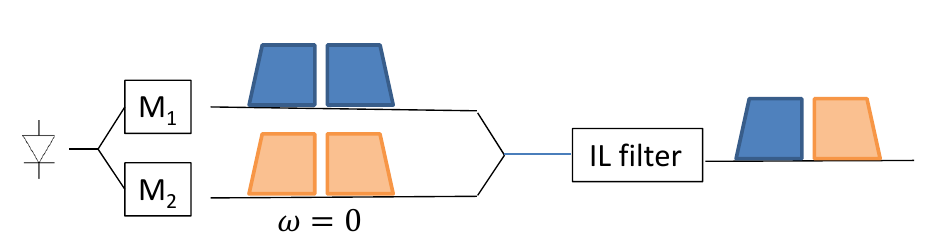}
	\caption{A different version of the same operating principle of Fig. \ref{OperPrinc}. Two single-quadrature modulators generate two real-valued signals, and a guard-band is inserted around $\omega=0$, so as to accommodate their separation at the receiver. An interleaving filter combines the upper sideband of one signal with the lower sideband of the other. The receiver scheme is identical to that of Fig. \ref{OperPrinc}.}\label{OperPrinc2}
\end{figure}
In that scheme the bandwidth reduction was obtained by offsetting the frequencies of the east-west channels with respect to the west-east channels, so that the information-carrying signal is always two-sided with respect to the transmission laser.

Here we propose an alternative approach to better exploit the transmitter bandwidth while keeping it equal to the receiver bandwidth, and without misaligning the frequencies of the east-west and west-east channels.

The operating principle of the proposed approach is illustrated in Fig. \ref{OperPrinc}. Instead of generating SSB signals as in Fig. \ref{KKbasic}, each modulator generates a two-sideband spectrum, where each sideband contains an independent data-carrying signal. A guard-band around the center frequency ($\omega=0$) must be programed into the system so as to allow separation between the two sidebands on the receiver side. The receiver starts by optically separating the two sidebands, which are to be processed independently and in parallel by means of the KK algorithm. The two reception processes make use of the same LO positioned at $\omega=0$. As compared to SSB transmission of Fig. \ref{KKbasic}, with this approach the same transmitter is used to transmit two channels instead of one, so that its bandwidth is optimally exploited. Of course, the principle of operation described in Fig. \ref{OperPrinc} can be implemented in a WDM environment characterized by a multiplicity of transmitters and receivers. In order to avoid confusion, in what follows we use the term WDM channel to refer to the signal generated by a single transmitter module.

\begin{figure}
	\centering\includegraphics[width=0.9\columnwidth]{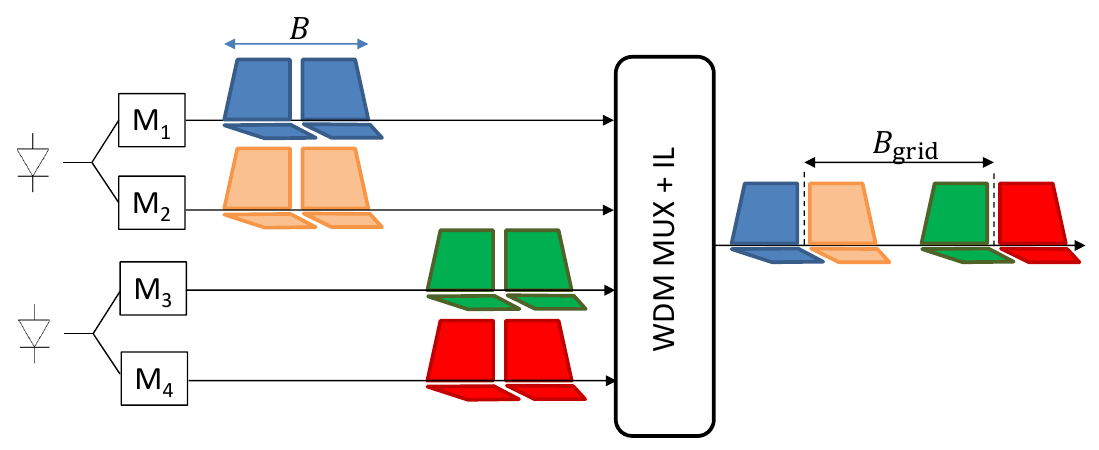}
	\caption{The transmitter of the TS-KK transceiver scheme. A transmission laser is shared by two single-quadrature modulators (either amplitude modulators or single Mach Zehnder modulators), each generating a real-valued signal of bandwidth $B$. The two signals are multiplexed first and then interleaved. The interleavers are off-set from the WDM grid (channel spacing $B_{\mathrm{grid}}$ in the figure) so as to suppress the high-frequency (low-frequency) sideband of the signals generated by the odd-indexed (even-indexed) modulators.}\label{SchemeTXpolmux}
\end{figure}

\begin{figure}
	\centering\includegraphics[width=0.95\columnwidth]{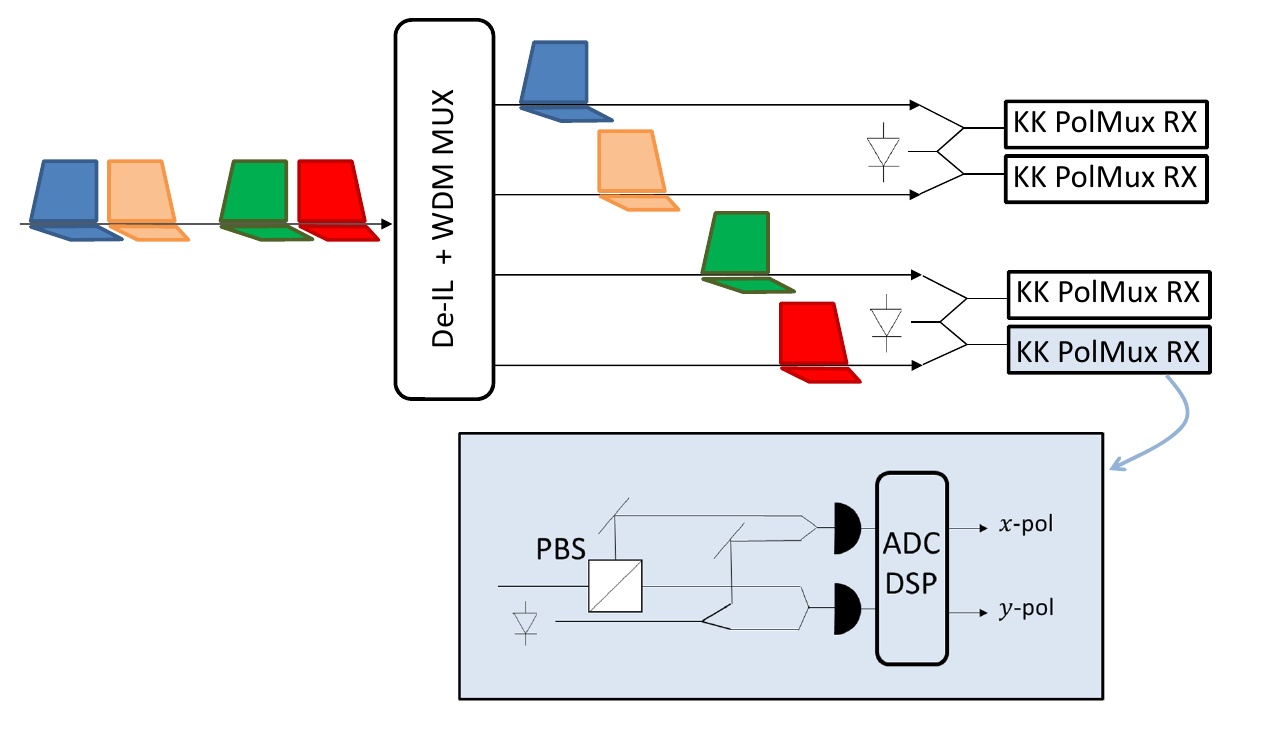}
	\caption{The receiver of the TS-KK transceiver scheme. The received WDM channels are de-interleaved first, so as to separate the low- and high-frequency sideband, and then de-multiplexed. Each pair of SSB signals extracted from the same WDM channel is finally received with two KK PolMux receivers sharing an LO. The KK PolMux receiver schematic is illustrated in the inset, while an extensive characterization can be found in \cite{KKPolMux}.  }\label{SchemeRXpolmux}
\end{figure}

A slightly different version of the same operating principle is illustrated in Fig. \ref{OperPrinc2}. Here, two single-quadrature modulators generate two real-valued signals, where, similarly to the previous case, a guard-band is inserted around $\omega=0$, so as to accommodate their separation at the receiver. An interleaving filter that combines the upper sideband of one signal with the lower sideband of the other, forms the WDM signal that is to be launched into the fiber. The schemes of Fig. \ref{OperPrinc} and \ref{OperPrinc2} use exactly the same receiver, but the comparison between the transmitters is interesting. The scheme of Fig. \ref{OperPrinc} relies on  an I/Q modulator, which consists of two single-quadrature modulators whose outputs need to be combined in quadrature with interferometric accuracy. In the case of Fig. \ref{OperPrinc2}, the two single-quadrature modulators are combined in frequency and no interferometric control is required. On the other hand, an interleaving filter needs to be included.  Of course, a single large interleaving filter can be shared by all WDM channels, as is illustrated in Fig. \ref{SchemeTXpolmux}, in the polarization-multiplexed case. The signals from even-indexed and odd-indexed transmitters must be multiplexed separately first, and then interleaved by means of two interleavers. At the receiver edge, a pair of interleavers is needed to de-interleave the SSB signals (of bandwidth $B/2$) prior to de-multiplexing.\footnote{We note that in principle the role of the interleavers and de-interleavers can also be played  by SSB filter pairs, or dichroic filters, incorporated into each transceiver.} This is illustrated in Fig. \ref{SchemeRXpolmux}.


Finally, as illustrated in Fig. \ref{SchemeRXpolmux}, each pair of SSB signals originated by paired transmitters is received by paired receivers that share a local oscillator. The sharing of the local oscillator between adjacent channels was previously proposed for balanced heterodyne detection in \cite{XLi}, and applied to KK receivers in \cite{ECOC17Plant}. 
Clearly,  the local oscillator grid used at the receiver end is in this case aligned with that used at the transmitter. 
{After de-multiplexing, each of the polarization-multiplexed channel sidebands  of bandwidth $B/2$ together with the local oscillator is polarization-split and detected by one photodiode per polarization. The two photocurrents are sampled by two ADCs of bandwidth $B$. 
A schematic of the KK polarization-multiplexed receiver is shown in the inset of Fig. \ref{SchemeRXpolmux}, while extensive descriptions can be found in \cite{KKECOC,KKPolMux,KKPolMuxVivian}. As mentioned earlier, the guard-band around each channel's central frequency is needed in order to allow the combining and separation of the two sidebands. The guard-band width is determined in conjunction with the filtering capability of available interleavers  and de-interleavers. When the guard-band is not sufficiently large, the residual spectral content of an imperfectly suppressed sideband falls on the wrong side of the LO, thus violating the minimum-phase condition, and thereby resulting in reconstruction errors.  

\begin{figure}
	\centering\includegraphics[width=0.9\columnwidth]{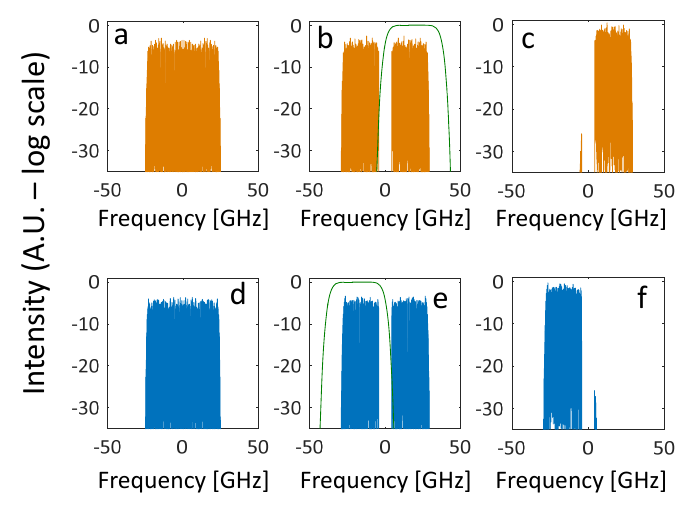}
	\caption{Generation of a two-sided signal, as illustrated in the transmitter scheme of Fig. \ref{SchemeTXpolmux}. The two spectral sides of two real-valued zero-mean 4ASK signals (a, d) are digitally shifted apart from each other (b, e). After digital-to-analog conversion, they are optically filtered with super-gaussian filters of fourth order (c, f).   }\label{Spectra}
\end{figure}

\subsection{Numerical validation of the two-sided Polarization-multiplexed KK Transceiver}
In this section we present a numerical validation of the TS-KK  transceiver scheme. We transmitted two single-sided zero-mean 4ASK signals over a fiber link consisting of five 100-km spans. To this end, we first generated two real-valued 4-ASK signals at 48 Gbaud, using a raised-cosine fundamental waveform with a roll-off factor of 0.05. We then introduced a frequency gap {$B_\mathrm{gap}$} by moving the positive and negative sides of the symmetric spectra apart from each other. This procedure is illustrated in Fig. \ref{Spectra}.  We assumed a WDM channel spacing of 80 GHz, and used filters with a forth-order super-gaussian shape, 3-dB bandwidth of 36 GHz, and 90\%--to--10\% roll-off bandwidth of 7 GHz to model the interleavers \cite{Interleavers} (solid curves in Figs. \ref{Spectra}b and \ref{Spectra}e). Good performance was achieved by using a frequency gap {$B_\mathrm{gap} = 8.6$ GHz}, while offsetting the interleaver grid by 18.8 GHz from the WDM grid. These settings yield good sideband suppression, as can be seen in Figs. \ref{Spectra}c and   \ref{Spectra}f, with no significant increase of the necessary transmitter bandwidth, {which in this case was increased from $B_\mathrm{net} = 48$ GHz to $B = 48 + 8.6 = 56.6 $ GHz}. At the receiver, the two sidebands of each WDM channel were separated with the same type of filters prior to adding the CW tone. After intensity detection, we used the KK algorithm to reconstruct the single-sideband complex signal, and compensated for chromatic dispersion, as well as for the polarization rotation caused by the fiber birefringence. Finally, the original real-valued 4ASK signal was  recovered by taking the real part of reconstructed and compensated signal, after digitally removing the frequency gap. All simulations were based on the use of the Manakov equation \cite{WaiMenyuk} and were performed without polarization mode dispersion. A frequency-independent random polarization was applied to the optical signal prior to reception in every run.

Figure \ref{BERpolmux}a shows the average BER in the linear operation regime as a function of the OSNR, for various intensities of the LO. The BER was evaluated on the basis of fifty independent simulation runs, each with  $2^{15}$ Grey-coded symbols, and averaged over the two signal's polarizations. Solid markers refer to the 4-ASK signal encoded in the high-frequency sideband, 
while empty markers refer to that encoded in the low-frequency sideband. The solid curve is the plot of the theoretical BER for 4-ASK.

In Fig. \ref{BERpolmux}b we study the performance of the TS-KK transceiver in the nonlinear operation regime. In this case, we transmitted five WDM channels and measured the BER of the central channel for the same LO powers used in Fig. \ref{BERpolmux}a.  As detailed in \cite{KKOptica}, when the LO is relatively weak ($4P_s$ in the figure), errors at large OSNR are mainly due to the reconstruction process, and hence the effect of the fiber nonlinearity is not visible. For large LO power levels and in the limit of large OSNR, where  reconstruction  errors are practically absent, the effect of nonlinearity becomes dominant. We note that in all cases considered here, the BER does not exceed the relevant threshold of $10^{-3}$.

The TS-KK transceiver is expected to achieve the performance of a coherent system with the same channel throughput and the same net bandwidth. 
In order to establish a quantitative comparison, we considered a 16-QAM coherent system operating at 48 Gbaud and 80 GHz channel spacing. The simulation results are shown in Fig. \ref{BERpolmux}b by stars, whereas the theoretical BER is still given by the solid curve. As is evident from the figure, for sufficiently large LO power the difference between coherent and TS-KK becomes negligible. 

\begin{figure}
	\centering\includegraphics[width=0.9\columnwidth]{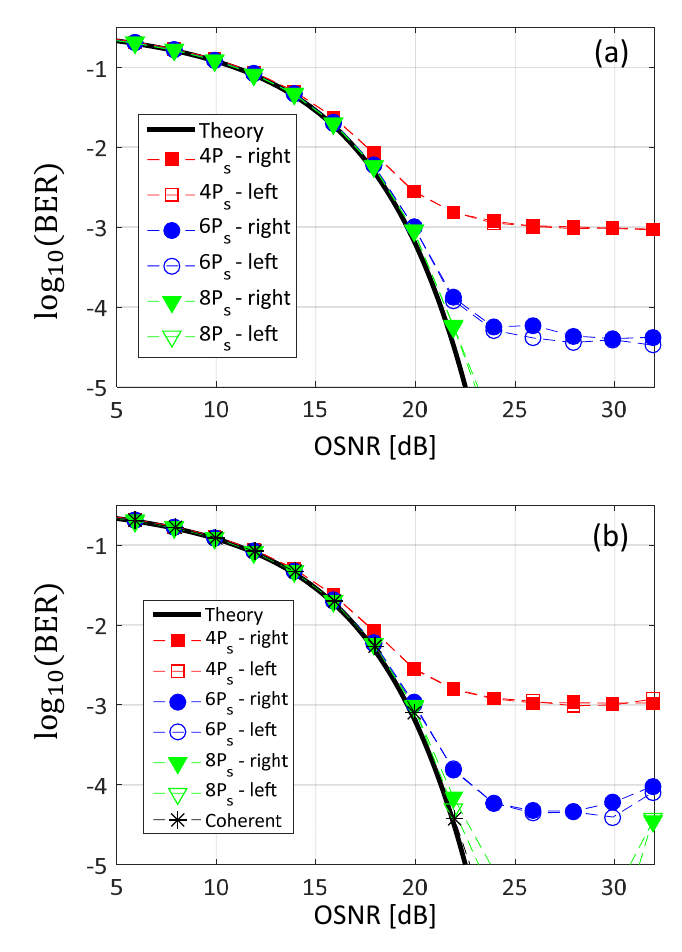}
	\caption{(a) Average BER in the linear operation regime versus OSNR, for the intensities of the LO shown in the legend.  Solid markers refer to the 4-ASK signal encoded in high-frequency sideband, while empty markers refer to that encoded in low-frequency sideband.  The solid curve is the plot of the theoretical BER for 4ASK. (b) Average BER in the nonlinear operation regime after 500 km propagation in a SMF versus OSNR, for the intensities of the LO shown in the legend. The stars show simulation results for a 16-QAM coherent system.}\label{BERpolmux}
\end{figure}
\subsection{Complexity comparison}
It is interesting to conclude this section with a comparison between the hardware complexity of the TS-KK scheme and the complexity of the two most relevant coherent communications schemes; intradyne\cite{Kikuchi} and balanced heterodyne.\footnote{Here, by balanced heterodyne (or simply heterodyne) we refer to the TS-KK scheme of Fig. \ref{SchemeRXpolmux}, where each of the two coupler-photodiode pairs in the KK PolMux RX block (inset) is replaced with a 50-50 bean splitter and a pair of balanced photodiodes.} The comparison is conducted for a polarization multiplexed system and for the case where the optical bandwidth of the transmitted WDM channel is $B$. As we have demonstrated above, the TS-KK scheme approaches the theoretical performance limit of coherent transmission. Hence, apart from the slightly lower spectral efficiency of TS-KK, the ultimate performance of all three schemes is identical. On the transmitter side, and restricting ourselves to typical implementations, all schemes require four optical modulators of electrical bandwidth $B$. In the coherent case, each modulator is responsible for modulating one quadrature, while the two quadratures in each polarization need to be combined with sub-wavelength accuracy. In the TS-KK transmitter, each modulator modulates one sideband channel, and since the sidebands are combined in frequency domain, no interferometric accuracy is needed from the combiner. On the other hand, the TS-KK receiver requires an optical interleaver filter, which does not exist in the coherent scheme. 

\begin{table}[htp]
	\caption{Device count at the receiver per channel (2 polarizations)}
	\begin{center}
		\begin{tabular}{cccccc}
			\hline
			\hline
			& \# LDs & \# PDs. &\# ADCs & \# OHs & Balancing \\
			\hline
			Intradyne & 1 & 8 & 4 & 2 & yes\\
			Heterodyne & 1 & 8 & 4 & 0 & yes \\
			TS-KK & 1 & 4 & 4 & 0 & no\\
			\hline
		\end{tabular}
	\end{center}
	Legend: LD: Laser diode; PD: Photodiode; ADC: Analog to digital converter;  OH: Optical hybrid. All devices have a bandwidth $B$.
	\label{Table1}
\end{table}

The comparison between the schemes on the receiver side is summarized in Table \ref{Table1}. All three schemes rely on the use of a LO, which requires tapping the light from one laser diode (also used for transmission). They also require the same number of ADCs, one per quadrature in the intradyne receiver and one per sideband channel in the case of heterodyne and TS-KK. The intradyne receiver requires two optical hybrids, one per polarization, whereas in the case of the heterodyne and TS-KK receivers, no optical hybrid is required. The number of photodiodes is eight in the case of the coherent receivers, as differential detection requires two balanced photodiodes per ADC. In the TS-KK receiver only four photodiodes are required --- one for each sideband channel.  On the other hand, the TS-KK receiver requires an extra interleaver, whose resolution requirements are higher than in the coherent case. We note however, that the cost of the extra interleavers required by the TS-KK scheme (at transmitter and receiver), is shared by all of the WDM channels.

\section{Conclusions}\label{Conclusions}
We have introduced two new approaches for implementing the KK field-reconstruction procedure in fiber-optic systems. We called them the KK-PAM and the TS-KK configurations. The KK-PAM scheme is similar to KK scheme in which the CW tone is combined with the data-carrying signal at the transmitter \cite{KKOFCpdp}. Its uniqueness is in the fact that it relies on the transmission of SSB PAM signals. The launched signal can be constructed digitally, by driving an I/Q modulator, or optically, in which case a single-quadrature modulator is used in combination with optical filtering. The TS-KK scheme relies on the availability of a LO at the receiver and its main advantage over KK-PAM (or the KK scheme of \cite{KKOFCpdp}), is in the fact that it can readily accommodate polarization multiplexing. Its construction is such that the LO can be extracted from the transmit laser that is present at the same location. The performance of the TS-KK is similar to that of typical coherent receivers, with the only drawback that a guard-band that allows the separation between sidebands needs to be included in the transmitted spectrum, thereby reducing the spectral efficiency. We have seen that with reasonable assumptions on the available components, the reduction of spectral efficiency is only within 15 percent. The advantage of the TS-KK scheme in comparison with coherent communications schemes, is in the absence of optical hybrids and in the fact that the outputs of the individual modulators at the transmitter need not be phase-controlled when combining them into a single fiber. In addition, the necessary number of photo-diodes is smaller by a factor of two, and no balancing is required. The disadvantage is the need of optical filtering, but the implied extra cost is shared by all WDM channels. Finally, it is worth pointing out that the TS-KK scheme is also suitable for medium-to-long haul transmissions, owing to the fact that the information-carrying signal is transmitted without a CW tone, which would be responsible for power inefficiency and extra nonlinear penalties. 


\section*{Ackowledgement} C. Antonelli and A. Mecozzi acknowledge financial support from the Italian Government under Cipe resolution n. 135 (Dec. 21, 2012), project INnovating City Planning through Information and Communication Technologies (INCIPICT). Mark Shtaif acknowledges the support of the Israel Science Foundation Grant No. 1401/16.

\bibliography{example.bib}
\end{document}